\documentclass[12pt]{article}
\usepackage{epsfig}
\begin{document}

\begin{center} {\bf \Large Axial anomaly: the modern status.}

\vspace{1.5cm} { B.L.IOFFE}

\vspace{3mm}
{\small \it Institute of Theoretical and Experimental Physics\\
B.Cheremushkinskaya 25, 117218 Moscow,Russia\\ ioffe@itep.ru}
\end{center}
\date{}



\newcommand{\be}{\begin{equation}}
\newcommand{\ee}{\end{equation}}

\def\la{\mathrel{\mathpalette\fun <}}
\def\ga{\mathrel{\mathpalette\fun >}}
\def\fun#1#2{\lower3.6pt\vbox{\baselineskip0pt\lineskip.9pt
\ialign{$\mathsurround=0pt#1\hfil##\hfil$\crcr#2\crcr\sim\crcr}}}

\def\Journal#1#2#3#4{{#1} {#2} (#4) #3 }
\def\NCA{{\em Nuovo Cimento} A}
\def\PHYS{{\em Physica}}
\def\NPA{{\em Nucl. Phys.} A}
\def\MATH{{\em J. Math. Phys.}}
\def\PRO{{\em Prog. Theor. Phys.}}
\def\NPB{{\em Nucl. Phys.} B}
\def\PLA{{\em Phys. Lett.} A}
\def\PLB{{\em Phys. Lett.} B}
\def\PLD{{\em Phys. Lett.} D}
\def\PL{{\em Phys. Lett.}}
\def\PRL{\em Phys. Rev. Lett.}
\def\PREV{\em Phys. Rev.}
\def\PREP{\em Phys. Rep.}
\def\PRA{{\em Phys. Rev.} A}
\def\PRD{{\em Phys. Rev.} D}
\def\PRC{{\em Phys. Rev.} C}
\def\PRB{{\em Phys. Rev.} B}
\def\ZPC{{\em Z. Phys.} C}
\def\ZPA{{\em Z. Phys.} A}
\def\ANNP{\em Ann. Phys. (N.Y.)}
\def\RMP{{\em Rev. Mod. Phys.}}
\def\CHEM{{\em J. Chem. Phys.}}
\def\INT{{\em Int. J. Mod. Phys.} E}

\vspace{1cm}

\begin{abstract}

The modern status  of the problem of axial anomaly in QED and QCD
is reviewed. Two methods of the derivation of the axial anomaly
are presented: 1) by splitting of coordinates in the expression
for the axial current and 2) by calculation of triangle diagrams,
where the anomaly arises from the surface terms in momentum space.
It is demonstrated, that the equivalent formulation of the anomaly
can be given, as a sum rule for the structure function in
dispersion representation  of three point function of AVV
interaction. It is argued, that such integral representation of
the anomaly has some advantages in the case of description of the
anomaly by contribution  of hadronic states in QCD. The validity
of the t'Hooft consistency condition is discussed. Few examples of
the physical application of the axial anomaly are given.
\end{abstract}


\vspace{5mm}

\section{Introduction}

The phenomenon  of anomaly plays an important role in quantum
field theory: in many cases it determines whether or not the
theory is selfconsistent  and can be realized in the physical
world and, therefore, allows one to select the acceptable physical
theories. In the given theory the anomalies often are related to
appearance  of new quantum numbers  (topological quantum numbers),
result in emerging of mass scale, determine the spectrum of
physical states. So, despite of its denomination, the anomaly is a
normal and significant attribute of any quantum field theory.

The term ``anomaly'' has the following meaning. Let the classical
action of the theory to obey some symmetry, i.e. it is invariant
under certain transformations. If this symmetry is violated by
account of quantum corrections, such a phenomenon is called   an
``anomaly''. (The reviews of anomalies are given in
\cite{Treiman}-\cite{Peskin}.) There are two types of anomalies --
internal and external. In the first case the gauge invariance of
the classical Lagrangian is destroyed  at the quantum level. The
theory becomes nonrenormalizable and cannot be considered as a
selfconsistent theory. The standard method to solve this problem
is the special choice of  fields in the Lagrangian in such a way,
that all internal anomalies  are cancelled. (The approach is used
in the Stanford Model of electroweak interaction -- it is the
Glashow, Illiopoulos, Maiani mechanism.) The external anomaly
corresponds to violation of symmetry of interaction with external
sources, not related to gauge symmetry of the theory. Just such
anomalies take place in QCD and are considered below. There are
two anomalies in QCD: the axial (chiral) anomaly and the scale
anomaly. Both are connected with singularities of the theory at
small distances (at large momenta) and with the necessity of
regularization: the regularization procedure, which respects to
the symmetry, does not exist and the symmetry is violated by the
anomaly. In QCD the evidence of anomalies came from perturbation
theory, but, in fact, their occurance follows from general
principles.


\section{The derivation the axial anomaly by
coordinate splitting}



The axial anomaly in QCD is very similar to those in massless QED.
For this reason let us first consider the latter. The equations of
motion of QED in the external electromagnetic field $A_{\mu}(x)$
have the form: \be i\gamma_{\mu} \frac{\partial\psi(x)}{\partial
x_{\mu}} = m\psi(x)-e\gamma_{\mu}A_{\mu}(x)\psi(x).\label{3.1}\ee
In massless QED, classically, i.e. without the account of
radiation corrections, the axial current $j_{\mu 5}(x)$ is
conserved like the vector current, \be
\partial_{\mu} j_{\mu 5} (x) = \partial_{\mu}j_{\mu}(x)
=0.\label{3.2}\ee However, it appears, that in quantum theory with
the account of radiation corrections, it is impossible to keep the
conservation of both currents -- vector and axial. The origin for
this comes from singular character of the currents. Vector and
axial currents  are composite operators built from local fermion
fields and the products of local operators are singular, when
their points coincide, as it is in the cases of $V$ and $A$
currents. In order to consider the problem correctly, define the
axial current by placing  two fermion fields at distinct points,
separated by the distance $\varepsilon$, and go to the limit
$\varepsilon \to 0$ in the final result, \be j_{\mu 5}
(x,\varepsilon) =\bar{\psi} \biggl ( x
+\frac{\varepsilon}{2}\biggr ) \gamma_{\mu}\gamma_5 \exp \biggl [
ie \int\limits^{x+\varepsilon/2}_{x-\varepsilon/2} dy_{\alpha}
A_{\alpha} (y) \biggr ] \psi \biggl (x-\frac{\varepsilon}{2}\biggr
).\label{3.3}\ee The exponential factor in (\ref{3.3}) is
introduced in order that the operator be locally gauge invariant.
The  divergence of axial current (\ref{3.3}) is equal (the
equation of motion (\ref{3.1}) is exploited and the first term in
the expansion in powers of $\varepsilon$ was retained): \be
\partial_{\mu}j_{\mu 5} (x,\varepsilon) = 2 im \bar{\psi}
\biggl (x+\frac{\varepsilon}{2}\biggr ) \gamma_5 \psi \biggl(x
-\frac{\varepsilon}{2}\biggr ) - ie\varepsilon_{\alpha} \bar{\psi}
\biggl (x+\frac{\varepsilon}{2}\biggr ) \gamma_{\mu}\gamma_5\psi
\biggl (x -\frac{\varepsilon}{2}\biggr ) F_{\alpha
\mu},\label{3.4}\ee where $F_{\alpha\mu}$ is the electromagnetic
field strength. For simplicity assume that $F_{\mu\nu}$=const.
Take the vacuum average of Eq.(\ref{3.4}). In the r.h.s. of
(\ref{3.4}) it can be used the expression for the electron
propagator in the constant external electromagnetic field. The
electron propagator \be S_{\alpha\beta}(x)=\langle 0\mid
T\{\psi_{\alpha}(x), \bar{\psi}_{\beta}(0)\} \mid 0 \rangle
\label{5}\ee satisfyes the equation: \be
[~i\gamma_{\mu}(\partial_{\mu} - ie A_{\mu}(x))-m~]
S(x)=i\delta^4(x).\label{6}\ee It is convenient to choose the
fixed point gauge for the electromagnetic field:
$x_{\mu}A_{\mu}(x)=0$. Then $a_{\mu}(x)$ is expressed through the
field strength tensor $F_{\mu\nu}$ by \be A_{\mu}(x) =\frac{1}{2}
x_{\nu} F_{\nu\mu}.\label{7}\ee The solution of Eq.(\ref{6}) up to
linear in $F_{\mu\nu}$ terms is equal \be S(x) =\frac{i}{2\pi^2}
\biggl [\frac{\not\!x}{x^4} +\frac{i}{2} \frac{m}{x^2}
+\frac{1}{16 x^2} e F_{\mu\nu} (\not\!x\sigma_{\mu\nu}
+\sigma_{\mu\nu}\not\!x)\biggr ],\label{8}\ee where
$\sigma_{\mu\nu} = (i/2)(\gamma_{\mu}\gamma_{\nu}
-\gamma_{\nu}\gamma_{\mu})$.
 The vacuum averaging
corresponds to account of the first order $e^2$ corrections. In
massless QED the first term in the r.h.s. of (\ref{3.4}) is absent
and we get \be \langle 0 \mid
\partial_{\mu}j_{\mu 5} \mid 0 \rangle =\frac{e^2}{4\pi^2}
F_{\alpha\mu} F_{\lambda \sigma} \varepsilon_{\beta\mu\lambda
\sigma}\frac{\varepsilon_{\alpha}\varepsilon_{\beta}}{\varepsilon^2},
~~~\varepsilon_{0123}=1. \label{3.5}\ee Since there is no specific
direction in space-time the limit $\varepsilon \to 0$  should be
taken symmetrically, \be \lim_{\varepsilon \to 0}
\frac{\varepsilon_{\alpha}\varepsilon_{\beta}}{\varepsilon^2}
=\frac{1}{4}\delta_{\alpha\beta}.\label{3.6}\ee The substitution
of (\ref{3.6}) into (\ref{3.5}) gives \be
\partial_{\mu}j_{\mu 5} =\frac{e^2}{8\pi^2} F_{\alpha \beta}
\tilde{F}_{\alpha \beta},\label{3.7}\ee where \be
\tilde{F}_{\alpha\beta} =\frac{1}{2} \varepsilon_{\alpha \beta
\lambda \sigma}F_{\lambda \sigma}\label{3.8}\ee  is the dual field
strength tensor. The symbol of vacuum averaging is omitted in
(\ref{3.7}), because in order $e^2$ Eq.(\ref{3.7}) can be
considered as an operator equation. The relation (\ref{3.7}) is
called the Adler-Bell-Jackiw anomaly \cite{Adler}-\cite{Jackiw}.


\section{The derivation of the axial anomaly by the
calculation of triangle diagrams}


In order to have the better understanding of the origin of the
anomaly let us consider the same problem in the momentum space. In
QED the  matrix element for the transition of the axial current
with momentum $q$ into two real or virtual photons with momenta
$p$ and $p^{\prime}$ is represented by the diagrams of Fig.1.
\begin{figure}[tb]
\hspace{44mm} \epsfig{file=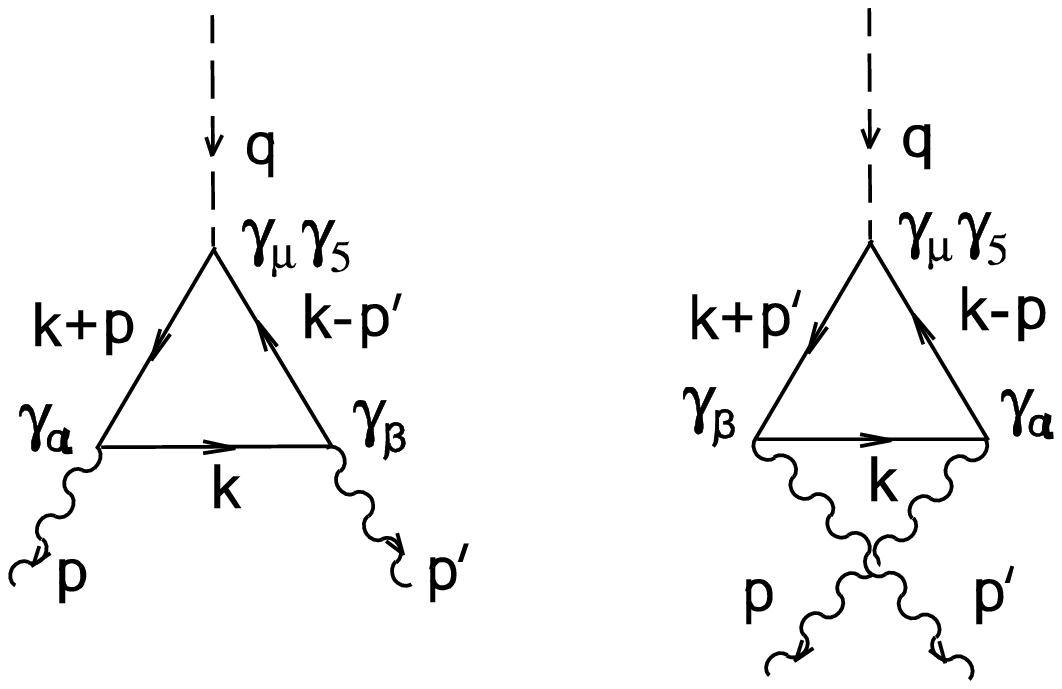, width=68mm}

\vspace{5mm}

{\bf Fig.1.}The diagrams, representing the vacuum expectation
value of axial current in the presence of external electromagnetic
field in QED, a) the direct diagram, b) the crossing diagram.

\end{figure}
 The matrix element is equal: \be
T_{\mu\alpha\beta}(p,p^{\prime}) =
\Gamma_{\mu\alpha\beta}(p,p^{\prime})
+\Gamma_{\mu\beta\alpha}(p^{\prime},p),\label{3.9}\ee \be
\Gamma_{\mu\alpha\beta}(p,p^{\prime}) = -e^2 \int \frac{d^4
k}{(2\pi)^4} Tr \biggl [ \gamma_{\mu}\gamma_5 (\not\!k +\not\!p
-m)^{-1}\gamma_{\alpha}(\not\!k -m)^{-1} \gamma_{
\beta}(\not\!k-\not\!p^{\prime} -m)^{-1}\biggr ].\label{3.10}\ee
Consider the divergence of the axial current $q_{\mu}
T_{\mu\alpha\beta}(p,p^{\prime})$, $q=p+p^{\prime}$. For
$q_{\mu}\Gamma_{\mu\alpha\beta}(p,p^{\prime})$ we can write (at
$m=0$): $$ q_{\mu}\Gamma_{\mu\alpha\beta} (p,p^{\prime}) =-e^2
\int \frac{d^4 k}{(2\pi)^4} Tr [~(\not\!p +\not\!k
+\not\!p^{\prime} -\not\!k) \gamma_5 (\not\!k+\not\!p)^{-1}
\gamma_{\alpha}\not\!k^{-1} \gamma_{\beta}(\not\!k
-\not\!p^{\prime})^{-1}~]=$$ \be = -e^2 \int \frac{d^4
k}{(2\pi)^4} Tr [~-\gamma_5\gamma_{\alpha}\not\!k^{-1}
\gamma_{\beta}(\not\!k -\not\!p^{\prime})^{-1}
-\gamma_5(\not\!k+\not\!p)^{-1}\gamma_{\alpha}\not\!k^{-1}
\gamma_{\beta}~]\label{3.11}\ee Each of the two terms in square
brackets in the right-hand side (r.h.s.) of (\ref{3.11}) after
integration of $k$ depends on only one 4-vector -- $p$ or
$p^{\prime}$. Each of these terms should be proportional to the
unit totally antisymmetric tensor
$\varepsilon_{\alpha\beta\gamma\beta}$ times the product of two
different vectors. Since we have only one vector at our disposal,
the result is zero. This fact looks to be in contradiction with
the anomaly relation (\ref{3.7}). However, we cannot trust in this
result. The arguments are the following. The integral (\ref{3.10})
is linearly divergent. In a linearly divergent integral it is
illegitimate to shift the integration variable: such shift may
result in appearance of the so-called ``surface terms''. So, if
the integration variable $k$ in (\ref{3.10}),(\ref{3.11}) would be
changed to $k+cp+dp^{\prime}$, where $c$ and $d$ are some numbers,
$q_{\mu}\Gamma_{\mu\alpha\beta}(p,p^{\prime})$ would not be zero.
The other argument against the calculation, performed in
(\ref{3.11}), is that $T_{\mu\alpha\beta}(p,p^{\prime})$ must
satisfy  the conditions of the conservation of vector current:
$p_{\alpha} T_{\mu\alpha\beta}(p,p^{\prime})=0$
$p^{\prime}_{\beta} T_{\mu\alpha\beta}(p,p^{\prime})=0$. The
calculations, performed using the same integration variable, as in
(\ref{3.11}) show, that these conditions are not fullfilled. The
question arises if it is possible to choose the integration
variable  in such a way, that $q_{\mu} T_{\mu\alpha\beta}=0$ and
simultaneously $p_{\alpha}T_{\mu\alpha\beta}=0$,
$p^{\prime}_{\beta} T_{\mu\alpha\beta}=0$. Following Ref.8,
consider $\Gamma_{\mu\alpha\beta}$, defined by (\ref{3.10}), where
the integration variable $k$ is shifted by an a arbitrary constant
vector $a_{\lambda}, k_{\lambda}\to k_{\lambda}+a_{\lambda}$. We
can write \be \Gamma_{\mu\alpha\beta}(p,p^{\prime};a) =
\Gamma_{\mu\alpha\beta}(p,p^{\prime}) +
\Delta_{\mu\alpha\beta}(p,p^{\prime},a)\label{3.12a}\ee \be
\Delta_{\mu\alpha\beta}(p,p^{\prime};a) =
\Gamma_{\mu\alpha\beta}(p,p^{\prime})_{k\to k+a} -
\Gamma_{\mu\alpha\beta}(p,p^{\prime})\label{3.13a}\ee where
$\Gamma_{\mu\alpha\beta}(p,p^{\prime})$ is given by (\ref{3.10})
and, therefore $q_{\mu}\Gamma_{\mu\alpha\beta}(p,p^{\prime})=0$
according to (\ref{3.11}).
$\Gamma_{\mu\alpha\beta}(p,p^{\prime})_{k\to k+a}$ is obtained
from (\ref{3.10}) by substituting $k_{\lambda}\to k_{\lambda} +
a_{\lambda}$. $\Delta_{\mu\alpha\beta}(p,p^{\prime};a)$ is the
surface term, the integral is convergent and its calculation gives
\cite{Jackiw}: \be \Delta_{\mu\alpha\beta} =-\frac{e^2}{8\pi^2}
\varepsilon_{\mu\alpha\beta\gamma}a_{\gamma}.\label{3.12}\ee
Generally, $a_{\lambda}$ is expressed in terms of two vectors,
involved in the problem -- $p$ and $p^{\prime}$,
$a_{\lambda}=(a+b) p_{\lambda} + b p^{\prime}_{\lambda}$.
Accounting the crossing diagram, we get: \be
T_{\mu\alpha\beta}(p,p^{\prime},a) =
T_{\mu\alpha\beta}(p,p^{\prime}) - \frac{e^2}{8\pi^2}
a\varepsilon_{\mu\alpha\beta\gamma}(p_{\gamma}-p^{\prime}_{\gamma}).\label{3.13}\ee
The matrix element of the divergence of the axial current appears
to be equal: \be q_{\mu} T_{\mu\alpha\beta}(p,p^{\prime};a)
=q_{\mu} T_{\mu\alpha\beta}(p,p^{\prime}) +\frac{e^2}{4\pi^2}
a\varepsilon_{\alpha\beta\gamma\sigma}p_{\gamma}p^{\prime}_{\sigma}.\label{3.14}\ee
As it was demonstrated above, the first term in r.h.s. of
(\ref{3.14}), vanishes in the limit of massless quarks (the
Sutherland-Veltman theorem \cite{Sutherland,Veltman}, see also
Ref.\cite{Jackiw}). As follows from (\ref{3.14}) in order to
ensure the conservation of the axial current it is necessary to
choose $a=0$. Such choice is just the repetition of the already
obtained result in Eq.(\ref{3.11}). Let us check now the
conservation of vector current. The direct calculation gives: \be
p_{\alpha} T_{\mu\alpha\beta}(p,p^{\prime};a) = \frac{e^2}{4\pi^2}
\varepsilon_{\mu\alpha\beta\gamma} p_{\alpha}
p^{\prime}_{\gamma}\biggl (1+\frac{a}{2}\biggr )\label{3.15}\ee
and the similar equality for $p^{\prime}_{\beta}
T_{\mu\alpha\beta} (p,p^{\prime};a)$. As follows  from
(\ref{3.15}) the conservation of vector current can be achieved,
if $a=-2$. That means, that it is impossible to have
simultaneously the conservation of vector and axial currents in
massless QED. Since we are sure, that the vector current is
conserved, otherwise the photon would be massive and all
electrodynamics would be ruined, we must choose $a=-2$. The
substitution  of $a=-2$ in (\ref{3.14}) gives back Eq.(\ref{3.7}).

Note, that the first method of the derivation of the anomaly -- by
the use of coordinate splitting in the expression for axial
current, is valid for constant external electromagnetic field,
since Eq.(6.273) corresponds to such case. The second method of
the derivation, based on consideration of the diagrams of Fig.1 is
much more general -- it is valid for arbitrary varying external
electromagnetic fields, including the emission of real or virtual
photons. In this case the anomaly condition has the form: \be
q_{\mu} T_{\mu\alpha\beta} (p,p^{\prime}) = \biggl [2 m
G(p,p^{\prime}) -\frac{e^2}{2\pi^2}\biggr ]
\varepsilon_{\alpha\beta\lambda\sigma} p_{\lambda}
p^{\prime}_{\sigma}.\label{3.16}\ee In (\ref{3.16}) the term,
proportional to electron mass is retained and $G(p,p^{\prime})$ is
defined by \be \langle p,\varepsilon_{\alpha};p^{\prime},
\varepsilon^{\prime}_{\beta} \mid \bar{\psi} \gamma_5 \psi\mid 0
\rangle = G(p,p^{\prime}) \varepsilon_{\alpha\beta\lambda\sigma}
p_{\lambda}p^{\prime}_{\sigma},\label{3.17}\ee where
$\varepsilon_{\alpha},\varepsilon^{\prime}_{\beta}$ are photon
polarizations.

The proof of the axial anomaly -- Eq.'s(\ref{3.7}),(\ref{3.16})
can be obtained also by other methods: by dimension reqularization
scheme, by Pauli -- Villars reqularization and by consideration of
functional integral \cite{Vergeles},\cite{Migdal},\cite{Fuijkawa}.
In the latter the axial anomaly arises due to noninvariance  of
the fermion measure in external gauge field at $\gamma_5$
transformations in functional integral.

Nevertheless,  the axial current is not conserved in massless QED,
there does exist a conserved, gauge invariant axial charge
\cite{Adler},\cite{Bell},\cite{Jackiw}. Define \be \tilde{j}_{\mu
5} =j_{\mu 5} -\frac{e^2}{4\pi^2} \tilde{F}_{\mu\nu}
A_{\nu}.\label{3.18}\ee The current $\tilde{j}_{\mu 5}$ is
conserved, but is not gauge invariant. However, the axial charge
\be Q_5 =\int d^3 x \tilde{j}_{0 5}(x)\label{3.19}\ee is gauge
invariant.

The axial anomaly in QED was considered till now in order of
$e^2$. It was shown, that there are no corrections to
Eq.(\ref{3.7}) in order $e^4$ \cite{Adler,AdlerSL} the argument is
that in this order all radiative corrections correspond to
insertion of photon line inside the triangle diagrams of Fig.1. If
the integration over the photon momentum is carried out after the
integration over the fermion loop, then the fermion loop integral
is convergent and there is no anomaly. (This argumentation was
supported by direct calculation \cite{AdlerSL}). In higher orders
of perturbation theory any insertions of photon lines and fermion
loops inside the triangle of Fig.1 diagrams do not give the
corrections to anomaly \cite{Adler},\cite{AdlerS.L.}. The
corrections to Adler-Bell-Jackiw anomaly arise from high order
diagrams like shown in Fig.2 \cite{Anselm}. The account of Fig.2
diagram results to renormalization of the anomaly term in
(\ref{3.7}),(\ref{3.16}) of the order $e^6$
\cite{Adler},\cite{AdlerS.L.},\cite{Anselm}.
\begin{figure}[tb]
\hspace{65mm} \epsfig{file=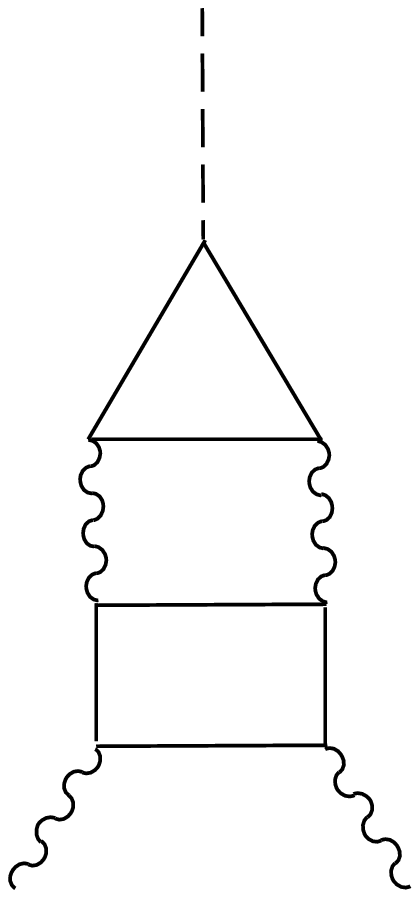, width=25mm}

\vspace{7mm}

\hspace{1.5cm}{\bf Fig.2.} The $e^6$ correction to
Adler-Bell-Jackiw anomaly in QED.

\end{figure}
In this order \cite{Anselm} \be \partial_{\mu}j_{\mu 5}
=\frac{\alpha}{2\pi} (F_{\mu\nu}\tilde{F}_{\mu\nu})_{ext}\biggl
(1-\frac{3}{4} \frac{\alpha^2}{\pi^2}\ln
\frac{\Lambda^2}{q^2}\biggr ),\label{3.22a}\ee where $\Lambda$ is
the ultraviolet cut-off, the axial vector vertex is renormalized
and the axial vector current, unlike the vector current, acquires
the anomalous dimension.

Turn now to QCD. Here $j_{\mu 5}$ can be identifyed with the
current of light quarks. In case of interaction with external
electromagnetic field if $j_{\mu 5}$ corresponds to the axial
current of one quark flavour with electric charge $e_q$ the
anomaly has the form of Eq,'s(\ref{3.7}),(\ref{3.16}) with the
only difference, that  the r.h.s. is multiplyed by $e^2_q N_c$,
where $N_c$ is the number of colours. In QCD there is also an
another possibility, where the external fields are gluonic fields.
In this case instead of (\ref{3.7}) we have: \be
\partial_{\mu} j_{\mu 5} =\frac{\alpha_s N_c}{4\pi} G^n_{\mu\nu}
\tilde{G}^n_{\mu\nu},\label{3.20}\ee where $G^n_{\mu\nu}$ is the
gluon field strength and $\tilde{G}^n_{\mu\nu}$ is its dual.
Eq.(\ref{3.20}) can be considered as an operator equation and the
fields $G^n_{\mu\nu},\tilde{G}^n_{\mu\nu}$ can be  represented by
virtual gluons. Note, that due to the same argumentation as in
case of radiation correction to the anomaly in QED, the
perturbative corrections  to (\ref{3.20}) start from $\alpha^3_s$.
Evidently, the flavour octet axial current \be j^i_{\mu 5} =\sum_q
\bar{\psi}_q \gamma_{\mu}\gamma_5 (\lambda^i/2)\psi_q,~~~i = 1,
...8\label{3.21}\ee is conserved in QCD. (Here $\lambda^i$ are
Gell-Mann $SU(3)$ matrices and the sum is performed over the
flavours  of light quarks, $q=u,d,s$.) Neglecting u,d,s quark
masses, we have instead of (\ref{3.20}): \be
\partial_{\mu}j^i_{\mu 5} =0.\label{3.22}\ee However, the anomaly
persists for singlet axial current \be j^{(0)}_{\mu 5} = \sum_q
\bar{\psi}_q \gamma_{\mu}\gamma_5 \psi_q,\label{3.23}\ee \be
\partial_{\mu}j^{(0)}_{\mu 5} = 3\frac{\alpha_s N_c}{4\pi}
G^n_{\mu\nu} \tilde{G}^n_{\mu\nu}\label{3.24}\ee From
(\ref{3.22}), (\ref{3.24}) it follows, that because of spontaneous
chiral symmetry breaking the octet of pseudoscalar mesons
$(\pi,K,\eta)$ is massless -- in the approximation $m_q\to 0$,
they are Goldstown bosons, while the singlet pseudoscalar meson --
the $\eta^{\prime}$ -- remains massive. Therefore, the occurance
of anomaly solve the so called $U(1)$ problem \cite{Weinberg}.
(The detailed exposition of this statement is given in
\cite{DyakonovDI}, see also \cite{Dyakonov},\cite{Leader} for
review.)


\section{The spectral representation of the three
point AVV function and the axial anomaly}

Return now to QED and consider the matrix element of the
transition of the axial current into two real or virtual photons,
i.e. the function $T_{\mu\alpha\beta}(p,p^{\prime})$ (\ref{3.9}),
described by the diagrams of Fig.3.2.1, where the internal lines
correspond to propagators of electrons. The general expression for
$T_{\mu\alpha\beta}(p,p^{\prime})$, which satisfyes the Bose
symmetry of two photons, reads
\cite{Adler},\cite{Bell},\cite{Eletsky}:
$$T_{\mu\alpha\beta}(p,p^{\prime})
=A_1(p,p^{\prime})S_{\mu\alpha\beta} - A_1(p^{\prime},p)
S^{\prime}_{\mu\alpha\beta} + A_2(p,p^{\prime})p_{\beta}
R_{\mu\alpha} -$$ \be - A_2(p^{\prime},p) p^{\prime}_{\alpha}
R_{\mu\beta} + A_3(p,p^{\prime}) p^{\prime}_{\beta} R_{\mu\alpha}
- A_3(p^{\prime},p) p_{\alpha} R_{\mu\beta},\label{3.25}\ee  where
\be R_{\mu\nu} =\varepsilon_{\mu\nu\rho\sigma}
p_{\rho}p^{\prime}_{\sigma},~~~~S_{\mu\alpha\beta}
=\varepsilon_{\mu\alpha\beta\sigma} p_{\sigma},
~~~~S^{\prime}_{\mu\alpha\beta}
=\varepsilon_{\mu\alpha\beta\sigma}p^{\prime}_{\sigma}.\label{3.26}\ee
The vector current conservation leads to
$$A_1(p,p^{\prime}) = (pp^{\prime}) A_2 (p,p^{\prime}) +
p^{\prime 2} A_3(p,p^{\prime})$$ \be A_1(p^{\prime},p) =
(pp^{\prime}) A_2 (p^{\prime},p) + p^2
A_3(p^{\prime},p)\label{3.27}\ee Using the identity \be
\delta_{\alpha\beta} \varepsilon_{\sigma\mu\nu\tau}
-\delta_{\alpha\sigma} \varepsilon_{\beta\mu\nu\tau}
+\delta_{\alpha\mu}\varepsilon_{\beta\sigma\nu\tau}
-\delta_{\alpha\nu}\varepsilon_{\beta\sigma\mu\nu}
+\delta_{\alpha\tau}\varepsilon_{\beta\sigma\mu\nu}=0,\label{3.28}\ee
we derive
$$ p_{\sigma} R_{\mu\nu} - p_{\mu} R_{\sigma\nu} + p_{\nu}
R_{\sigma\mu} + (pp^{\prime}) S_{\sigma\mu\nu} - p^2
S^{\prime}_{\sigma\mu\nu} =0$$ \be p^{\prime}_{\sigma} R_{\mu\nu}
- p^{\prime}_{\mu} R_{\sigma\nu} + p^{\prime}_{\nu} R_{\sigma\mu}
+ (pp^{\prime}) S^{\prime}_{\sigma\mu\nu} + p^{\prime 2}
S_{\sigma\mu\nu} =0.\label{3.29}\ee The Lorenz structures
$S_{\sigma\mu\nu}$ and $S^{\prime}_{\sigma\mu\nu}$ are retained in
(\ref{3.25}) in order to avoid kinematical singularities
\cite{Eletsky}. Let us put $p^2=p^{\prime 2} \leq 0$. Using
(\ref{3.27}) and the identities (\ref{3.29})
$T_{\mu\alpha\beta}(q,p,p^{\prime})$ can be expressed in terms of
two functions -- $F_1(q^2,p^2)$ and $F_2(q^2,p^2)$
\cite{Horejsi,Bass}:
$$ T_{\mu\alpha\beta}(p,p^{\prime}) = F_1(q^2,p^2)
q_{\mu}\varepsilon_{\alpha\beta\rho\sigma}
p_{\rho}p^{\prime}_{\sigma} - $$ \be -\frac{1}{2}
F_2(q^2,p^2)\biggl
[\varepsilon_{\mu\alpha\beta\sigma}(p-p^{\prime})_{\sigma}
-\frac{p_{\alpha}}{p^2} \varepsilon_{\mu\beta\rho\sigma}
p_{\rho}p^{\prime}_{\sigma} + \frac{p^{\prime}_{\beta}}{p^2}
\varepsilon_{\mu\alpha\rho\sigma}
p_{\rho}p^{\prime}_{\sigma}\biggr ].\label{3.30}\ee If
$p^2\not=0$, the form factors $F_1(q^2,p^2)=-A_2$ and
$F_2(q^2,p^2)=2A_1$ are free of kinematical singularities
\cite{Bass}. Consider now the divergence \be q_{\mu}
T_{\mu\alpha\beta}(p,p^{\prime}) = [~F_2(q^2,p^2) +q^2
F_1(q^2,p^2)~]
\varepsilon_{\alpha\beta\rho\sigma}p_{\rho}p^{\prime}_{\sigma}.\label{3.31}\ee
The substitution in the l.h.s. of (\ref{3.31}) of the anomaly
condition (\ref{3.16}) gives the sum rule \cite{Horejsi} \be
F_2(q^2,p^2)+q^2F_1(q^2,p^2) = 2m G(q^2,p^2) -
\frac{e^2}{2\pi^2}.\label{3.32}\ee The functions
$F_1(q^2,p^2),F_2(q^2,p^2)$ and $G(q^2,p^2)$ can be represented by
the unsubtracted dispersion relations in $q^2$: \be f_i(q^2,p^2)
=\frac{1}{\pi} \int\limits^{\infty}_{4m^2} \frac{Im
f_i(t,p^2)}{t-q^2} dt,~~~~f_i=F_1,F_2,G.\label{3.33}\ee The direct
calculation of $Im~F_1(q^2,p^2)$ gives
\cite{Horejsi},\cite{Frishman} $$ Im~F_1(q^2,p^2)
=-\frac{e^2}{2\pi} \frac{2p^2}{q^2} \left\{ \frac{q^2
+2p^2}{(q^2-4p^2)^2} \biggl ( 1-\frac{4m^2}{q^2} \biggr )^{1/2} +
\frac{2p^2(q^2-2p^2)}{(q^2)^{1/2}(q^2-4p^2)^{5/2}}\times\right.$$
\be\times \left. \biggl [ \frac{q^2-p^2}{q^2-2p^2}
+m^2\frac{q^2-4p^2}{2(p^2)^2}\biggr ] \ln \frac{q^2-2p^2
-[(q^2-4m^2)(q^2-4p^2)]^{1/2}}{q^2-2p^2
+[(q^2-4m^2)(q^2-4p^2)]^{1/2}}\right\}.\label{41}\ee At $p^2=0$
\be Im~F_1(q^2,0) = \frac{e^2}{\pi} \frac{m^2}{q^4} \ln
\frac{1+\sqrt{1-4m^2/q^2}}{1-\sqrt{1-4m^2/q^2}},\label{42}\ee and
at large $q^2$ \be Im~F_1(q^2,p) \approx -\frac{e^2}{2\pi} \cdot
\frac{2p^2}{q^4} \biggl [1+\frac{m^2}{p^2}\ln
\frac{m^2}{q^2}\biggr ]\label{43}\ee
 For imaginary parts of $F_1,F_2,G$
we have the relation: \be Im F_2 (q^2,p^2) + q^2 Im F_1 (q^2,p^2)
=2m Im G(q^2,p^2).\label{3.34}\ee As follows from (\ref{43}) and
(\ref{3.34}) $F_2(q^2,p^2)$ and $G(q^2,p^2)$ are decreasing as
$1/q^2$ at $q^2\to \infty$. Therefore, the nonsubtracted
dispersion relation (\ref{3.33}) are legitimate. From
(\ref{3.32})-(\ref{3.34}) the sum rule \be
\int\limits^{\infty}_{4m^2} Im F_1 (t,p^2)dt
=\frac{e^2}{2\pi}\label{3.35}\ee follows. The sum rule
(\ref{3.35}) has been verified explicitly  by Frishman et al
\cite{Frishman} for $p^2 <0,m=0$, by
Ho$\check{\mbox{r}}$ej$\check{\mbox{s}}$i \cite{Horejsi} at
$p^2=p^{\prime 2} < 0$ and by Veretin  and Teryaev \cite{Veretin}
in general case, $p^2\not=p^{\prime 2}$.

Consider now the transition of axial current into two real photons
in QCD with one flavour of unit charge. Instead of (\ref{3.35}) we
have  \be \int\limits^{\infty}_{4m^2_q} Im F_1(t,0)dt =2\alpha
N_c.\label{3.36}\ee  $F_2(q^2,p^2)$ should vanish at
$p^2\Rightarrow 0$ in order that
$T_{\mu\alpha\beta}(p,p^{\prime})$ have no pole there, which would
correspond to massless hadronic state in $J^{PC}=1^{--}$ channel.
 In the limit of massless quarks, $m_q=0$, the r.h.s. of
(\ref{3.32})  is given by the anomaly and in QCD we have \be
F_1(q^2,0)_{m^2_q=0} = -\frac{2\alpha N_c}{\pi}
\frac{1}{q^2},\label{3.37}\ee \be T_{\mu\alpha\beta}
(p,p^{\prime}) =- \frac{2\alpha}{\pi} N_c \frac{q_{\mu}}{q^2}
\varepsilon_{\alpha\beta\lambda\sigma}p_{\lambda}p^{\prime}_{\sigma}.\label{3.38}
\ee  The imaginary part of  $F_1(q^2,0)$ at $m^2_q=0$ is
proportional to $\delta(q^2)$ \cite{Dolgov}:  \be
Im~F_1(q^2,0)_{m^2_q=0}=2\alpha N_c \delta(q^2)\label{3.39a}\ee
and the sum rule (\ref{3.36}) is saturated by the contribution of
zero-mass state. It is interesting to look how the limit $m^2_q\to
0,q^2\to 0$ proceeds. At $m_q\not=0~Im~F_1(q^2,0)$ is equal
\cite{Dolgov} \be Im~F_1(q^2,0)=4\alpha N_c \frac{m^2}{q^4} \ln
\frac{1+\sqrt{1-4m^2_q/q^2}}{1-\sqrt{1-4m^2_q/q^2}}
\label{3.40a}\ee and in the limit $m^2_q\to 0,q^2 \to 0$ indeed we
get (\ref{3.39a}). The most interesting case is, when the current
$j_{\mu 5}$ is equal to the third component of the isovector
current \be j^{(3)}_{\mu 5} =\bar{u} \gamma_{\mu}\gamma_5 u -
\bar{d}\gamma_{\mu}\gamma_5d.\label{3.39}\ee Then at
$p^2=p^{\prime 2}=0, m^2_u=m^2_d=0$ \cite{Dolgov}: \be
T_{\mu\alpha\beta} (p,p^{\prime}) =-\frac{2\alpha}{\pi} N_c
\frac{q_{\mu}}{q^2} (e^2_u -e^2_d) \varepsilon_{\alpha
\beta\lambda\sigma} p_{\lambda}p^{\prime}_{\sigma}.\label{3.40}\ee
The amplitude $T_{\mu\alpha\beta}$ (\ref{3.40}) corresponds to the
transition of isovector axial current into two photons.
Eq.(\ref{3.40}) is consistent with the fact that the transition
proceeds through virtual $\pi^0$ and $\pi^0$ is massless at
$m_q=0$ (the pole in the amplitude at $q^2=0$) \cite{Dolgov}. Then
the process is described by the diagram like Fig.3.
\begin{figure}[tb]
\hspace{60mm} \epsfig{file=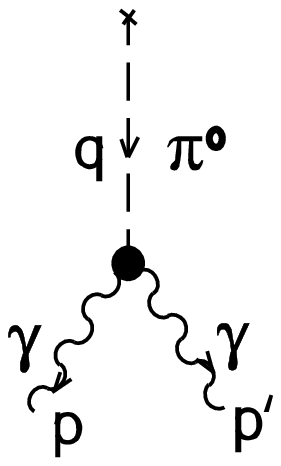, width=25mm}

 {\bf Fig.3.} The diagram describing the
transition of of isovector axial current (marked by cross) into
two photons through virtual $\pi^0$.

\end{figure}
The use of relation $\langle 0 \mid j^{(3)}_{\mu 5} \mid
\pi^0\rangle =\sqrt{2}if_{\pi} q_{\mu}$ determines the amplitude
of $\pi^0\to 2\gamma$ decay \be M(\pi^0 \to 2\gamma)
=A\varepsilon_{\alpha \beta \lambda \sigma}
\varepsilon_{1\alpha}\varepsilon_{2\beta}p_{1\lambda}p_{2\sigma},\label{3.41}\ee
where $\varepsilon_{1\alpha}$ and $\varepsilon_{2\beta}$ are the
polarizations of the first and the second photons. From
(\ref{3.40}) the constant $A$ is found to be \be
A=\frac{2\alpha}{\pi} \frac{1}{\sqrt{2}f_{\pi}}\label{3.42}\ee and
the $\pi^0\to 2\gamma$ decay rate is equal \be \Gamma(\pi^0 \to
2\gamma) =\frac{\alpha^2}{32 \pi^3}
\frac{m^3_{\pi}}{f^2_{\pi}}.\label{3.43}\ee Eq.(\ref{3.43}) gives
the theoretical value of $\pi^2 \to 2\gamma$ decay width
$\Gamma(\pi^0\to 2\gamma)_{theor}=7.7$ eV in very good agreement
with experimental value $\Gamma(\pi^0\to 2\gamma)_{\exp}=7.8+ 0.6$
eV \cite{PDG}. (The accuracy of theoretical value (\ref{3.43}) is
5-7\%. The higher accuracy of theoretical prediction was achieved
in \cite{Goity} in the framework of CET and $1/N_c$ expansion.)

 Despite of the fact, that the axial
anomaly results in appearance of massless $\pi^0$ in the
transition of isovector axial current into two photons and
predicts well the $\pi^0$ decay rate, it is incorrect to say that
the existence of massless pseudoscalar Goldstone bosons (at
$m_q=0$) are caused by the anomaly. The reasons are the following.
$Im~ F_1(q^2,p^2)$ has $\delta(q^2)$ singularity at $p^2=0$.
According to the Chiral Effective Theory (CET) it is expected that
the same singularity persist in the case of $p^2\not= 0$ -- the
diagram of Fig.3. contributes in this case as well. However, the
examination of $Im~ F_1(q^2,p^2)$, Eq.(\ref{41}) shows, that $Im
F_1(q^2, p^2)$ is a regular  function of $q^2$ near $q^2=0$ at
$p^2\not= 0$. The sum rules (\ref{3.35}),(\ref{3.36}) are
satisfyed by the triangle diagram contribution at $p^2\not =0$.
Therefore, it is fulfilled the statement of CET, that the
transition of $j^{isov}_{\mu 5}$ into $2\gamma$'s with $p^2\not=0$
is described by the diagram of Fig.3.2.3 with the same $\pi^0\cdot
2\gamma$ coupling constant as at $p^2=0$, but only in the sense of
integrals (\ref{3.35}),(\ref{3.36}), not locally.  (It is assumed
that $\mid p^2 \mid$ is less than CET characteristic mass scale).
Consider now the transition of 8-th component of octet axial
current \be j^{(8)}_{\mu 5} =\frac{1}{\sqrt{6}}  (
\bar{u}\gamma_{\mu}\gamma_5 u +\bar{d} \gamma_{\mu} \gamma_5 d -
2\bar{s} \gamma_{\mu}\gamma_5 s)\label{3.44}\ee into two real
photons at $m_u=m_d=m_s=0$. The amplitude $F_1(q^2,0)$ has a pole
at $q^2=0$, which can be attributed to $\eta$-meson. The $\eta\to
2\gamma$ decay width is given by the relation, analogous to
(\ref{3.43}) \be \Gamma(\eta\to 2\gamma) =\frac{\alpha^2}{32\pi^3}
\frac{1}{3} \frac{m^3_{\eta}}{f^2_{\eta}}.\label{3.45}\ee However,
(\ref{3.45}) strongly disagrees with experiment: $\Gamma(\eta \to
2\gamma)_{theor} =0.13$ keV (at $f_{\eta}=150$ Mev) in comparison
with $\Gamma(\eta\to 2\gamma)_{\exp} = 0.510\pm 0.026$ keV \cite
{PDG}.  The possible explanation of this discrepancy is  strong
nonperturbative interactions like instantons, which persist in
pseudoscalar channel (see \cite{Geshkenbein}). The
$\eta\eta^{\prime}$ mixing remarkably increases $\Gamma(\eta \to
2\gamma)$. Another discrepancy arises, if we consider the the
transition $j^{(0)}_{\mu 5} \to 2\gamma$, where $j^{(0)}_{\mu 5}$
is the singlet axial current: \be j^{(0)}_{\mu 5}
=\frac{1}{\sqrt{3}}(\bar{u}\gamma_{\mu}\gamma_5 u +\bar{d}
\gamma_{\mu}\gamma_5 d + \bar{s} \gamma_{\mu}\gamma_5
s).\label{3.46}\ee Since at $m_u=m_d=m_s=0$ there are the poles at
$q^2=0$ in $F_1(q^2,0)$ for each quark flavour, the transition
amplitude $T^{(0)}_{\mu\alpha\beta}(q,p,p^{\prime})$ has a pole at
$q^2=0$. The corresponding pseudoscalar meson is $\eta^{\prime}$.
But $\eta^{\prime}$ is not a Goldstone boson -- it is massive! The
possible explanation is  the important role of instantons in
$\eta^{\prime}\to 2\gamma$ decay \cite{Dorokhov} and that the
contribution of the diagram similar to Fig.3 (with virtual gluons
instead of virtual photons) and, may be, the ladder (or parquet)
of box diagrams is of importance here \cite{Veretin}. (We do not
touch the theoretical determination of $\eta\to 2\gamma$ and
$\eta^{\prime}\to 2 \gamma$ decay rates by using additional
hypothesis, besides the anomaly condition -- see \cite{Shore} and
references therein.)

Turn now back to Eq.'s(\ref{3.35}),(\ref{3.36}). These equations
are equivalent to anomaly conditions. The integrals in the l.h.s.
of these equations are convergent. ($Im~F_1(q^2,p^2)_{q^2\to
\infty}\sim 1/q^4$). This means, that with such interpretation the
anomaly arises from finite domain of $q^2$. Eq.(\ref{3.35}) can be
rewritten in another form\cite{Veretin}: \be \lim_{q^2\to\infty }
q^2 \pi F_1(q^2,p^2) =\frac{e^2}{2\pi}.\label{3.47}\ee This form
returns us to initial interpretation of the anomaly, as
corresponding to the domain of infinitely large $q^2$. So, it is
possible to speak about the double face of the anomaly: from one
point of view it corresponds to large $q^2$, from the other -- its
origin is connected with finite $q^2$. As is clear from the
discussion above, both points of view are correct. These two
possibilities of interpretation of the anomaly are interconnected
by analyticity of the corresponding amplitudes.

 t'Hooft suggested the hypothesis, that the
singularities of amplitudes, calculated in QCD on the level of
quarks and gluons shall be reproduced on the level of hadrons (the
so called t'Hooft consistency condition \cite{t'Hooft}). Of
course, if it is possible to prove, that such singularity cannot
be smashed out by perturbative  and nonperturtbative corrections,
this statement is correct and, even more, it is trivial. But, as a
rule, no such proof can be given. In the presented above examples
of the realization of axial anomaly (with the exception of
$\pi^0\to 2\gamma$ decay) t'Hooft conjecture  was not realized.
Much better chances are for the duality conditions, like
Eq.(\ref{3.36}), when the QCD amplitude, integrated over some
duality interval, gives the same result, as the corresponding
hadronic amplitude integrated over the same duality interval (the
so called quark-hadron duality).

In QCD the case, when one of the photons  in Fig.3.2.1 is soft, is
of special interest \cite{Vainshtein}. If the momentum of the soft
photon is $p^{\prime}_{\beta}$ and its polarization is
$\varepsilon^{\prime}_{\beta}$, then, restricting ourselves by the
linear terms in $p^{\prime}_{\beta}$, the amplitude
$T_{\mu\alpha\beta} \varepsilon^{\prime}_{\beta}$ can be
represented in terms of two structure functions:
$$T_{\mu\alpha\beta} \varepsilon^{\prime}_{\beta} = w_T (q^2) (
-q^2\tilde{f}_{\alpha\mu} +q_{\alpha}
q_{\sigma}\tilde{f}_{\sigma\mu} -
q_{\mu}q_{\sigma}\tilde{f}_{\sigma\alpha})+ $$ \be + w_L(q^2)
q_{\mu}q_{\sigma} \tilde{f}_{\sigma\alpha},\label{3.50a}\ee where
\be \tilde{f}_{\mu\nu} =\frac{1}{2}
\varepsilon_{\mu\nu\lambda\sigma} (p^{\prime}_{\lambda}
\varepsilon^{\prime}_{\sigma}
-p^{\prime}_{\sigma}\varepsilon^{\prime}_{\lambda}).\label{3.51a}\ee
The first structure is transversal with respect to axial current
momentum $q_{\mu}$, while the second is longitudinal. From the
triangle diagram the relation \cite{Veretin},\cite{Achasov}: \be
w_L(q^2) =2 w_T(q^2) \label{3.52a}\ee follows. The anomaly
condition gives for massless quark: \be w_L(q^2) =2
\frac{\alpha}{\pi} N_c \frac{1}{q^2}.\label{3.53a}\ee Because of
(\ref{3.52a}) this condition determines also the transverse
structure function. According to the Adler-Bardeen
nonrenormalization theorem, there are no perturbative corrections
to the triangle diagram. But, as was demonstrated  in
\cite{Vainshtein} there are nonperturbative corrections, which at
large $q^2$ can be expressed through OPE series in terms of vacuum
condensates, induced by external electromagnetic field. In terms
of OPE Eq.(\ref{3.53a}) represents the contribution of the
dimension 2 operator $\tilde{F}_{\mu\nu}$. The next in dimension
vacuum condensate is the quark condensate magnetic susceptibility
$\chi(d=3)$, defined by \be \langle 0\mid \bar{q} \sigma_{\mu\nu}
q \mid 0 \rangle_F = e_q F_{\mu\nu} \langle 0 \mid \bar{q}q \mid 0
\rangle \chi, \label{64}\ee which was introduced in
\cite{Ioffe},\cite{Smilga}. The index $F$ in (\ref{64}) means that
the vacuum expectation value is taken in the presence of the
constant weak electromagnetic field $F_{\mu\nu}$. It was assumed
in Ref.\cite{Vainshtein}, that only the lowest hadronic state --
the pion contributes to the anomaly and $q^2$ in the denominator
was substituted by $q^2-m^2_{\pi}$. Then the expansion in
$m^2_{\pi}$ in the first order and identification of this term
with dimension 3 term of OPE (the proportional to $1/q^4$ term in
$w_L$)  allows one to find the quark condensate magnetic
susceptibility \be \chi =-\frac{N_c}{2\pi^2 f^2_{\pi}} =
-8.9~\mbox{GeV}^{-2}\label{65}\ee The value of $\chi$, determined
from QCD sum rules is equal \cite{Balitsky} \be \chi_{1~GeV} =
-4.4 \pm 0.4~\mbox{GeV}^{-2}\label{66}\ee The disagreement of
(\ref{65}) and (\ref{66}) cannot be considered as a strong
discrepancy. $\chi$ has anomalous dimension $(d=-16/27)$. No
$\alpha_s$-corrections were accounted  in (\ref{65}), therefore it
is not clear to what scale the value (\ref{65}) refers. The
saturation by pion contribution can be valid at low scale, where
$\alpha_s$-corrections are large. The contribution of excited
state are of the same order as the pion contribution -- there are
no small parameter there. For all these reasons (\ref{65}) can be
considered as the order of magnitude estimation of $\chi$ and  the
comparison of (\ref{65}) and (\ref{66}) is merely an argument in
favour of the used approach.


\section{The axial anomaly and the scattering of
polarized electron (muon) on polarized gluon}


 Consider the scattering of longitudinally polarized
electron (muon) on longitudinally polarized gluon. The first
moment of the forward scattering amplitude is proportional to the
diagonal matrix element  \be \langle g_{polar}\mid j_{\mu 5}\mid
g_{polar}\rangle,\label{3.48}\ee where the gluons are on mass
shell. The corresponding Feynman diagrams are the same as in Fig.1
with the only difference, that the wavy lines represent now the
polarized gluons and the lower vertices are the vertices of
quark-gluon interaction. Put $q=0, p=-p^{\prime}, p^2 <0$. It is
convenient to use the light-cone kinematics, where $p_0=p_+ +
p_-/2$, $p_z=p_+ -p_-/2$, $p^2=2p_+ p_- < 0$ and work in the
infinitely fast moving system along the $z$-direction. The matrix
element is given by: $$ \Gamma_{\mu}(p) =2ig^2 N_f~ Tr\biggl
(\frac{\lambda^n}{2}\biggr )^2 \int \frac{d^4 k}{(2\pi)^4} Tr \{~
\not\!{\varepsilon}^* (\not\!{k}+m)\gamma_{\mu}\gamma_5 (\not\!{k}
+m)\not\!{\varepsilon} (-\not\!p +\not\!k+m)~\}$$ \be \times
\frac{1}{(k^2-m^2+i\varepsilon)^2}\frac{1}{[(p-k)^2
-m^2+i\varepsilon]}. \label{3.49}\ee Here $N_f$ is the number of
flavours, $\lambda^n, n=1, ...8$ are the  Gell-Mann $SU(3)$ matrix
in colour space, $m$ are the quark masses, which are assumed to be
equal for  all flavours, $\varepsilon_{\mu}$ is the gluon
polarization vector, \be \varepsilon_{\mu} =\frac{1}{\sqrt{2}}
(0,1, i,0) \label{3.50}\ee for gluon helicity $+1$. The
contribution of the crossing diagram Fig.3.2.1b is equal to the
direct one and is accounted in (\ref{3.49}) by the factor 2. The
calculation of (\ref{3.49}) is performed using the dimensional
regularization in $ n\not=4$ dimensions. According to the
t'Hooft-Veltman recipe (see, e.g.\cite{Collins}) it is assumed,
that $\gamma_5$ is anticommuting with $\gamma_{\mu}$ at
$\mu=0,1,2,3,$ and is commuting with $\gamma_{\mu}$  at $\mu \not=
0,1,2,3$. After integration over $k_-$ it was found for the
component $\Gamma_{5+}$ \cite{Carlitz}: $$ \Gamma_{5+} =
-\frac{\alpha_s N_fp_+}{\pi^2} \int\limits^1_0 dx \int
\frac{d^{n-2} k_T}{[k^2_T +m^2 +P^2x (1-x)]^2} \left \{k^2_T(1-2x)
- m^2 -\right.$$ \be \left. -2\biggl ( \frac{n-4}{n-2}\biggr )
k^2_T (1-x) \right \},\label{3.51}\ee where $P^2=-p^2$. In the
integration over $k_-$ it was enough to take the residue at the
pole of the last propagator in (\ref{3.49}), what results in the
integration domain in $k_+$: $0 < k_+ < p_+$, and allowed to put
$k_+ =xp_+$. The last term in (\ref{3.51}) arised from $n-4$
regulator dimensions and is proportional to $\hat{k}^2$, where
$\hat{k}$ is the projection of $k$ into these dimensions. The
azimuthal average gives $\hat{k}^2 = k^2_T(n-4)/(n-2)$.

The first term in the curly brackets in (\ref{3.51}) vanishes
after integration over $x$. (In fact, the equal to zero result
after integration over $x$  is multiplyed by divergent integral
over $k_T$. So, strictly speaking,  this term is uncertain. This
problem will be discussed later.) After integration over $k_T$,
using the rules of dimensional regularization and going to $n=4$,
we get \cite{Achasov}: \be \Gamma_{5^+} = - \frac{\alpha_s N_f
p_+}{\pi} \left\{ 1 -\int\limits^1_0 \frac{2m^2(1-x) dx}{m^2 +P^2
x (1-x)}\right\}.\label{3.52}\ee The first term in the  r.h.s. of
(\ref{3.52}) arises from the last term in (\ref{3.51}) and is of
ultraviolet origin. As was stressed by Gribov \cite{Gribov} and in
Ref.\cite{Carlitz} it cannot  be addressed to any definite set of
guark-gluon configurations and is a result of collective effects
in  QCD vacuum. In other words, this term can be considered as a
local probe of gluon helicity.

The magnitude of $\Gamma_{s^+}$ strongly depends on the ratio
$m^2/P^2$. At $m^2/P^2 \ll 1$ \be \Gamma_{5^+} =-\frac{\alpha_s
N_f p_+}{\pi}.\label{3.53}\ee In the opposite case, $m^2/P^2 \gg
1$ the second term in r.h.s. of (\ref{3.52}) almost entirely
cancels the first one and approximately \be \Gamma_{5^+} \approx
0.\label{3.54}\ee The real physical situation corresponds to the
first case. Gluons do not exist as free particles, they are
confined in hadrons and their  virtualities  are of order of
inverse confinement radius squared, $P^2 \sim R^{-2}_c \gg m^2$.

Turn now to a more detailed discussion of the contribution of the
first term in the curly brackets in (\ref{3.51}). At fixed $k_T$
this contribution is zero, because the integration over $x$: the
denominator is symmetric under interchange  $x\leftrightarrow
(1-x)$, while the numerator  is antisymmetric under such
interchange. For the same reason the contribution of this term
vanishes at dimensional regularization at $n\not=4$. However, the
domain of low $k^2_T\leq P^2$ contributes to the integral over
$k^2_T$ here. In this domain we cannot be sure, that the integrand
in the first term in (\ref{3.51}) has the same form as it is
presented there. If this form is different -- we know nothing
about it -- and if it is not  symmetric under interchange
$x\rightarrow (1-x)$, then nonvanishing infrared contribution to
$\Gamma_{5^+}$ can arise from this term \cite{Bass}. Consider the
simple model with infrared cut-off in $k^2_{\perp}$, \be k^2_T >
M^2 (x, P^2),\label{3.55}\ee where $M^2(x,P^2)$ is some function
of $x,P^2$. In the first term of (\ref{3.49}) the integral over
$k^2_T$ can be written as the integral in the limits $(0,\infty)$,
which vanishes after integration over $x$, as before, minus the
integral in the limits $(0,M^2)$. As  a result, neglecting the
term, proportional to $m^2$, instead of (\ref{3.53}) we get
\cite{Bass}: \be \Gamma_{5^+} = -\frac{\alpha_s N_f p_+}{\pi}
\left\{ 1 - \int\limits^1_0 dx (1-2x) [\ln r(x)
-r(x)]\right\},\label{3.56}\ee where \be r(x)
=\frac{x(1-x)P^2}{x(1-x)P^2 +M^2(x,P^2)}.\label{3.57}\ee
Eq.(\ref{3.56}) demonstrates, that the matrix element $\langle
g_{polar}\mid j_{\mu 5} \mid g_{polar}\rangle $ is not entirely
contributed by ultraviolet domain, connected with the anomaly, but
can get the contribution from infrared region.

Note, that in the parton model $\Gamma_{5+}$ is related to the
part of hadron spin carried by gluons in polarized hadron: \be
(\triangle g^h_1)_{gl} = \int\limits^1_0 g^h_{1,gl} (x) dx
=(\Gamma_{5+}/2p_+) [~g_{gl+} -g_{gl-}~].\label{3.58}\ee Here
$g_{gl+}$ and $g_{gl-}$ are the  numbers of gluons in  hadron,
with helicities $+1$ and $-1$ correspondingly, $g_{1,gl}^h(x)$  is
the contribution of gluons to the structure function $g^h_1(x)$.
For polarized proton it was found
\cite{Efremov},\cite{Altarelli},\cite{Anselmino} \be (\Delta
g_1)_{gl} = -\frac{\alpha_s N_f}{2\pi} (g_{gl+} -
g_{gl-}).\label{3.59}\ee


\section{Summary}

 The analysis of the  axial anomaly was performed not
only by the standard methods, but also by the use of the less well
known method -- the method of the spectral representation of the
three point AVV amplitude
\cite{Horejsi},\cite{Frishman},\cite{Veretin}. It was shown, that
the latter has some advantages in comparison with the formers.
E.g. it allows one to describe not only the decay of $\pi^0$ into
two real photons, as was done earlier \cite{Dolgov}, but also into
two virtual ones. It was argued that in cases of octet and singlet
axial currents in QCD nonperturbative effects are important. The
t'Hooft consistency condition was discussed and it was
demonstrated, that it is not universal: the nonperturbative
effects can spoil condition.

\vspace{8mm}

\centerline{\bf \large Acknowledgement}

\bigskip

\noindent I am thankful to A.S.Gorsky for useful remarks. This
work was supported  in part by US CRDF Cooperative Grant Program,
Project RUP2-2621-MO-04, RFBR grant 06-02-16905-a and the funds
from EC to the project ``Study of Strongly Interacting Matter''
under contract No.R113-CT-2004-506078.

\end{document}